\documentclass[preprint,showpacs,preprintnumbers,amsmath,amssymb]{revtex4}
\usepackage{booktabs}
\usepackage{mathrsfs}
\usepackage{epsfig}
\usepackage{graphicx}
\usepackage{dcolumn}
\usepackage{bm}
\usepackage{amsmath}
\usepackage{slashed}       

\let\jnfont=\rm
\def\NPB#1,{{\jnfont Nucl.\ Phys.\ B }{\bf #1},}
\def\PLB#1,{{\jnfont Phys.\ Lett.\ B }{\bf #1},}
\def\EPJC#1,{{\jnfont Eur.\ Phys.\ Jour.\ C }{\bf #1},}
\def\PRD#1,{{\jnfont Phys.\ Rev.\ D }{\bf #1},}
\def\PRL#1,{{\jnfont Phys.\ Rev.\ Lett.\ }{\bf #1},}
\def\MPLA#1,{{\jnfont Mod.\ Phys.\ Lett.\ A }{\bf #1},}
\def\JPG#1,{{\jnfont J.\ Phys.\ G}{\bf #1},}
\def\CTP#1,{{\jnfont Commun.\ Theor.\ Phys.\ }{\bf #1},}
\def\ZPC#1,{{\jnfont Z.\ Phys.\ C }{\bf #1},}
\def\JHEP#1,{{\jnfont JHEP \ }{\bf #1},}
\def\lsim{\raise0.3ex\hbox{$<$\kern-0.75em\raise-1.1ex\hbox{$\sim$}}}
\def\gsim{\raise0.3ex\hbox{$>$\kern-0.75em\raise-1.1ex\hbox{$\sim$}}}

\begin{document}

\title{Status of low energy SUSY models confronted with \\ the LHC 125 GeV Higgs data}

\author{Junjie Cao$^{1,2}$, Zhaoxia Heng$^1$, Jin Min Yang$^3$, Jingya Zhu$^3$}

\affiliation{
  $^1$  Department of Physics,
        Henan Normal University, Xinxiang 453007, China \\
  $^2$ Center for High Energy Physics, Peking University,
       Beijing 100871, China \\
  $^3$ State Key Laboratory of Theoretical Physics,
      Institute of Theoretical Physics, Academia Sinica, Beijing 100190, China
      \vspace{1cm}}

\begin{abstract}
Confronted with the LHC data of a Higgs boson around 125 GeV,
different models of low energy SUSY show different behaviors:
some are favored, some are marginally survived and some are strongly
disfavored or excluded. In this note we update our previous scan over
the parameter space of various low energy SUSY models by considering
the latest experimental limits like the LHCb data for $B_s\to \mu^+\mu^-$
and the XENON 100 (2012) data for dark matter-neucleon scattering.
Then we confront
the predicted properties of the SM-like Higgs boson in each model
with the combined 7 TeV and 8 TeV Higgs search data of the LHC.
For a SM-like Higgs boson around 125 GeV, we have the following observations:
(i) The most favored model is the NMSSM, whose predictions about the Higgs boson
can naturally (without any fine tuning) agree with the experimental data
at $1\sigma$ level, better than the SM;
(ii) The MSSM can fit the LHC data quite well but suffer from some extent of
fine tuning;
(iii) The nMSSM is excluded at $3\sigma$ level after considering all the
available Higgs data;
(iv) The CMSSM is quite disfavored since it is hard to give a 125 GeV Higgs
boson mass and at the same time cannot enhance the di-photon signal rate.
\end{abstract}

\maketitle

\section{Introduction}
The LHC experiments have just reported compelling evidence for a
Higgs boson around 125 GeV: combining the 7 TeV and 8 TeV
data the ATLAS and CMS collaborations have separately obtained
a $5\sigma$ local significance  \cite{ATLAS,CMS,recent-LHC}. This observation is
corroborated by the Tevatron results which showed a $2.5\sigma$ excess
in the range 115-135 GeV  \cite{tevatron-higgs}.
So far although the limited data show consistency with the SM prediction,
the best-fit results nevertheless favor
a non-standard Higgs   \cite{best-fit,h-fit1,best-fit1}.
In fact, ever since this 125 GeV particle was
first hinted last year, some analyses of its property
were quickly performed in new physics models, such as
the low energy SUSY models \cite{125-susy,125-cao,carena,125-GMSB,cao-cmssm,125-cmssm,125-mssm}
and some non-SUSY models  \cite{125-other}.
For the SUSY models the following
observations were obtained: (i) The minimal gauge mediation SUSY
breaking model (GMSB) and anomaly mediation
SUSY breaking model (AMSB) cannot predict a SM-like Higgs boson as heavy as 125 GeV without
incurring severe fine-tuning  \cite{125-GMSB}; while the constrained minimal
supersymmetric standard model (CMSSM) and the minimal supergravity model (mSUGRA) can
marginally accommodate a 125 GeV Higgs boson \cite{cao-cmssm,125-cmssm};
(ii) The minimal supersymmetric standard model (MSSM), the nearly minimal
supersymmetric model (nMSSM) and
the next-to-minimal  supersymmetric model (NMSSM) can readily predict a 125 GeV
Higgs boson \cite{125-mssm}, but the MSSM suffers from some fine tuning \cite{tuning1}
and the nMSSM severely suppresses the di-photon signal rate \cite{hinvd,di-photon3}.

Now with the 8 TeV data, the statistics is much larger than last year's 7 TeV
data so that more accurate Higgs information can be extracted, which enable us to testify
different new physics models. Although one can envisage roughly that some
models are (dis)favored by the data, it is necessary to confront directly the model
predictions with the experimental data and show in detail the extent
of their compatibility. For this purpose we consider four low energy SUSY models, namely the
CMSSM, MSSM, NMSSM and nMSSM, and compare the predictions of each model, such
as the Higgs signal rates, the decay branching ratios and the couplings, with the latest data.
These predictions will be calculated in the parameter space
obtained from our previous works  \cite{125-cao,cao-cmssm,hinvd},
which satisfy various experimental constraints such as the precision
electroweak data, the B-decays, the muon $g-2$ and the dark matter
relic density in the $2\sigma$ range. Since some experimental limits
have just been updated, e.g., the
latest LHCb data for $B_s\to \mu^+\mu^-$ \cite{LHCb-Bdecay}
and the XENON 100 (2012) data for dark
matter-neucleon scattering \cite{XENON2012},
we will first update the scan and then give the
predictions in the allowed parameter space.
Note that since the GMSB and AMSB must have scalar top partners (stops) as heavy
as about 10 TeV in order
to give a 125 GeV Higgs boson \cite{125-GMSB}, they can no longer be called low
energy SUSY and thus will not be analyzed in this work.

This work is organized as follows.
In Sec. II, we recapitulate the four models and their Higgs sector features.
In Sec. III, we calculate the Higgs couplings,
the decay branching ratios and the signal rates in comparison to
the experimental data.
Finally, we draw our conclusions in Sec. IV.

\section{Low energy SUSY models and their features in Higgs sector}
{\bf MSSM}: As the most economical realization of SUSY in particle physics, the
MSSM has the superpotential given by $W_F+\mu \hat H_u\cdot\hat H_d$
where $W_F= Y_u  \hat{Q}\cdot\hat{H}_u  \hat{U}
    -Y_d \hat{Q}\cdot\hat{H}_d \hat{D}
    -Y_e \hat{L}\cdot\hat{H}_d \hat{E}$
with  $\hat{H}_u$ and $\hat{H}_d$ denoting
the Higgs doublet superfields, $\hat{Q}$, $\hat{U}$ and $\hat{D}$
denoting the squark superfields and $\hat{L}$ and $\hat{E}$ denoting the slepton
superfields \cite{mssm}.
Since there are two Higgs doublets in model construction, the MSSM predicts
five physical Higgs bosons, among which two are CP-even
($h$ and $H$), one is CP-odd ($A$) and the other two are a pair of charged
ones ($H^\pm$). At tree level this Higgs sector is determined by two parameters,
usually taken as  $m_A$ and $\tan \beta \equiv \frac{v_u}{v_d}$ with
$v_u$ and $v_d$ representing the vacuum expectation values of the two Higgs doublets.
In general, the lightest Higgs boson $h$ is SM-like, and for moderate $\tan \beta$ and large $m_A$,
its mass is given by\cite{carena}
\begin{equation}\label{mh}
 m^2_{h}  \simeq M^2_Z\cos^2 2\beta +
  \frac{3m^4_t}{4\pi^2v^2} \ln\frac{M^2_{S}}{m^2_t} +
\frac{3m^4_t}{4\pi^2v^2}\frac{X^2_t}{M_{S}^2} \left( 1 -
\frac{X^2_t}{12M^2_{S}}\right),
\end{equation}
where the first term on the right side is the tree level mass and the last two terms are the
dominant corrections from top-stop sector with $v=174 {\rm ~GeV}$, $X_t \equiv A_t - \mu \cot \beta$
denoting the scalar top mixing and $M_{S}$ representing the average stop mass scale defined by
$M_{S}=\sqrt{m_{\tilde{t}_1}m_{\tilde{t}_2}}$.  As indicated by numerous studies, in order to drive $h$
as heavy as 125 GeV, $M_S$ or $X_t$ must be large, which will induce somewhat fine tuning problem.
The di-photon signal rate for such a boson at the LHC is usually less than its SM prediction except
for a small fraction of the MSSM parameter space characterized by
light $\tilde{\tau}$ and large $\mu \tan \beta$\cite{carena}, while the $pp\to h\to ZZ^* \to 4\ell$ and
$pp\to h\to WW^* \to 2\ell+2\nu$  signal rates can
never get enhanced in the MSSM \cite{125-cao}.

{\bf CMSSM}: The CMSSM is same as the MSSM except for the assumption of the boundary
condition for its soft mass parameters. At the boundary (usually the GUT scale), the
soft parameters are assumed to be $m_0$ for scalar masses, $m_{1/2}$ for gaugino
masses and $A_0$ for trilinear couplings. As a result, the parameter
space of the CMSSM is rather limited in comparison with that of the MSSM. For example,
it was found  that $m_h$ is upper bound by about 124 GeV (126 GeV)  before (after)
considering its theoretical uncertainty if one takes into account
experiments constraints on the model, especially considers those from the muon $g-2$ and the
decay $B_s\to \mu^+\mu^-$ \cite{cao-cmssm}. This is because the SUSY explanation of
the muon $g-2$ at $2\sigma$ level requires relatively light slepton and gaugino masses, which
means not-so-heavy $m_0$ and $m_{1/2}$. In this case the only way to enhance $m_h$ is through large $X_t$,
which in return will raise the branching ratio of $B_s\to \mu^+\mu^-$ \cite{cao-cmssm}.
About this model, one should remind that both the di-photon and $Z Z^\ast$
signal rates cannot get enhanced.

{\bf NMSSM}: The NMSSM  extends the MSSM by introducing a gauge singlet
superfield $\hat{S}$ with the $Z_3$-invariant superpotential given by
$ W_F + \lambda\hat{H_u} \cdot \hat{H_d} \hat{S}+\kappa \hat{S^3}/3$ \cite{NMSSM}.
As a result, the NMSSM predicts one more CP-even Higgs
boson and one more CP-odd Higgs boson, and the $\mu$-term is dynamic generated
once the singlet scalar $S$ develops a vev. Corresponding to the superpotential,
new soft breaking terms like $\tilde m_S^2|S|^2$, $A_\lambda \lambda SH_u\cdot H_d$ and
$A_\kappa \kappa S^3/3$ appear, which will complicate the dependence of the Higgs mass matrices
on the model parameters. It has been shown that at tree level the SM-like Higgs boson
mass square receives an additional term $\lambda^2v^2\sin^2(2\beta)$, which, together with the mixing effect
among the doublet and singlet Higgs fields, may make the large radiative correction unnecessary, thus ameliorates
the fine tuning suffered by the MSSM \cite{tuning}. Moreover,
since  the SM-like Higgs boson in this model has the singlet component,
its coupling to $b\bar{b}$ can be suppressed and so is its total decay width.
This is helpful to enhance the branching ratio of $h\to \gamma\gamma$
and its related di-photon signal rate at the LHC \cite{125-cao}. Similar situation
applies to the $pp\to h\to ZZ^* \to 4\ell$ and
$pp\to h\to WW^* \to 2\ell+2\nu$ signals \cite{125-cao}.

In the limit of vanishing $\lambda$ and $\kappa$ (but for fixed $\mu$), the singlet field decouples
from the doublet Higgs sector in the NMSSM and the MSSM phenomenology is recovered. This motives
us that in order to get the Higgs properties significantly different from the MSSM prediction,
on should consider large $\lambda$ case. For example, on condition that $\lambda > ( m_Z/v \simeq 0.53)$,
the tree level mass of the SM-like Higgs boson in the NMSSM is maximized at low $\tan \beta$, instead of
large $\tan \beta$ in the MSSM.

{\bf nMSSM}: The nMSSM is same as the NMSSM except that
the cubic singlet term $\kappa \hat{S}^3$ in its superpotential is
replaced by a tadpole term $\xi_F M_n^2 \hat{S}$  \cite{xnMSSM,nMSSM}.
Clearly this potential has no discrete symmetry and thus free of the
domain wall problem \cite{xnMSSM}. The unique feature of this model
is the lightest neutralino as the lightest supersymmetric particle
(LSP) and the dark matter candidate is singlino dominated and must be light.
This is because the singlino mass term in the neutralino mass matrix vanishes,
and the LSP gets its mass only through the mixing of the singlino with higgsinos.
For such a dark matter, it must annihilate through exchanging a resonant light
CP-odd Higgs boson to get the correct relic density. As a result, although the sturcture
of the Higgs sector in the nMSSM of is quite similar to that of the NMSSM,
the SM-like Higgs boson tends to decay dominantly into light neutralinos
or other light Higgs bosons \cite{nMSSM} so that its total width
enlarges greatly. This will greatly suppress the di-photon
signal rate as well as the $WW^\ast$ and $ZZ^\ast$ signal rates.

\section{Higgs properties in confrontation with the LHC data}
First we update our scan \cite{125-cao,cao-cmssm,hinvd} by
considering the latest experimental limits and enlarging the scan ranges
for $M_{Q_3}$, $M_{U_3}$ and $|A_t|$.
In our scan we require SUSY to explain the muon $g-2$ anomaly
at $2\sigma$ level and at the same time satisfy the following experimental
constraints:
(i) the experimental bounds on sparticle masses;
(ii) the Higgs searches from the LEP and LHC experiments \cite{Atlas-h-2011,CMS-h-2011};
(iii) the $2\sigma$ limits from the precision electroweak data
and various B physics observables like the branching ratios of $B\to X_s \gamma$ and
$B_s \to \mu^+ \mu^-$ \cite{LHCb-Bdecay};
(iv) the dark matter constraints
including its relic density (the $ 2 \sigma$ range given by the WMAP) and
the direct search limits from XENON100 (2012) experiment at $90\%$ confidence
level.

In our calculations we use the packages NMSSMTools-3.2.0 \cite{NMSSMTools} and
HiggsBounds-3.8.0 \cite{HiggsBounds}.
For  the cross section of dark matter-nucleon scattering,
we use our own code by setting $f_{Ts}=0.025$ \cite{cao-dark}.
Note that we check the MSSM
results by using the code FeynHiggs  \cite{FeynHiggs}.
We found good agreement between
NMSSMTools and  FeynHiggs when $m_h$ lies within $125 \pm 2$ GeV.
For the CMSSM, we first use the NMSPEC in NMSSMTools
to run the soft breaking parameters from the GUT scale down to the weak scale,
then compute the couplings, branching ratios and signal rates of the SM-like Higgs
boson with the FeynHiggs \cite{cao-cmssm}.

The details of our scan are described in  \cite{125-cao,cao-cmssm,nMSSM}.
Here we only list the scan ranges of the NMSSM parameters
(the scan ranges for $M_{Q_3}$, $M_{U_3}$ and $|A_t|$ are enlarged
compared with our previous studies):
\begin{eqnarray}\label{NMSSM-scan}
&& 0.53 <\lambda\leq 0.7,~ 0<\kappa \leq 0.7, ~90{\rm ~GeV}\leq M_A\leq 1 {\rm ~TeV},
~|A_{\kappa}|\leq 1{\rm ~TeV},
\nonumber\\
&& 100{\rm ~GeV}\leq M_{Q_3},M_{U_3}
\leq 2 {\rm ~TeV} ,~~|A_{t}|\leq 5 {\rm ~TeV}, \nonumber\\
&& 1\leq\tan\beta \leq 60,  ~~100{\rm
~GeV}\leq \mu,m_{\tilde{l}}\leq 1 {\rm ~TeV},~~50{\rm~GeV}\leq M_1\leq 500 {\rm ~GeV},
\end{eqnarray}
Here we require a large $\lambda$ because for a small $\lambda$
the NMSSM is very similar to the MSSM.
In the following we only keep the surviving
samples which predict $ 123 {\rm GeV} \leq m_h \leq 127 {\rm GeV}$
and pay special attention to
the 'golden samples' which predict $m_h$ in the $1 \sigma$ best-fit
range $125.5 \pm 0.54$ GeV \cite{best-fit}.

\begin{figure}[t]
\centering
\includegraphics[width=10.8cm]{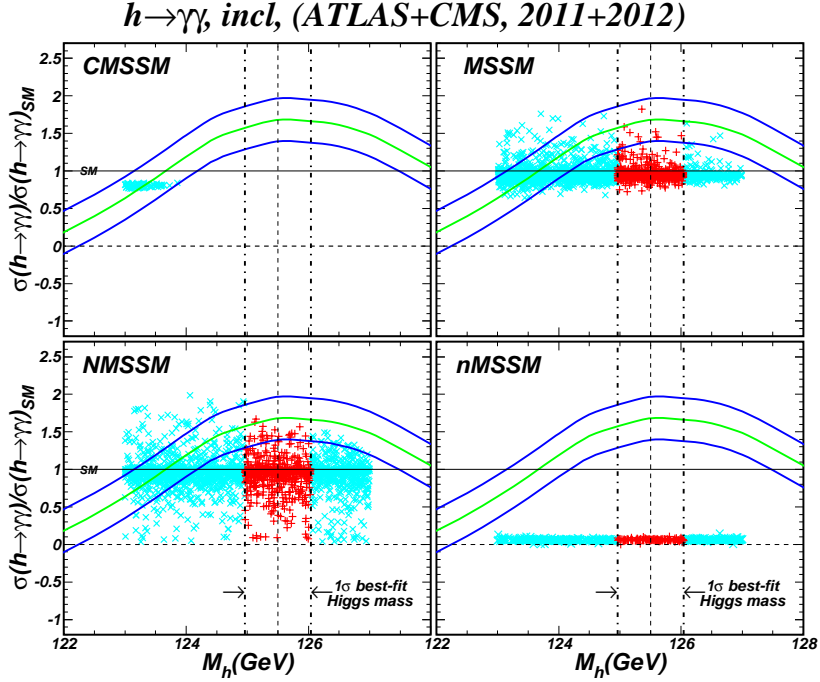}
\vspace*{-0.5cm}
\caption{The scatter plots of the samples surviving all the constraints
and predicting $m_h=125\pm 2$ GeV,
projected on the plane of the di-photon signal rate versus $m_h$.
The curves denote the central value and the $1\sigma$ region of the LHC data.
The samples denoted by '+' (red)
predict a SM-like Higgs boson in the best-fit range $125.5 \pm 0.54$ GeV
(called 'golden samples' in the text).}
\label{fig1}
\end{figure}
\begin{figure}[t]
\centering
\includegraphics[width=10.8cm]{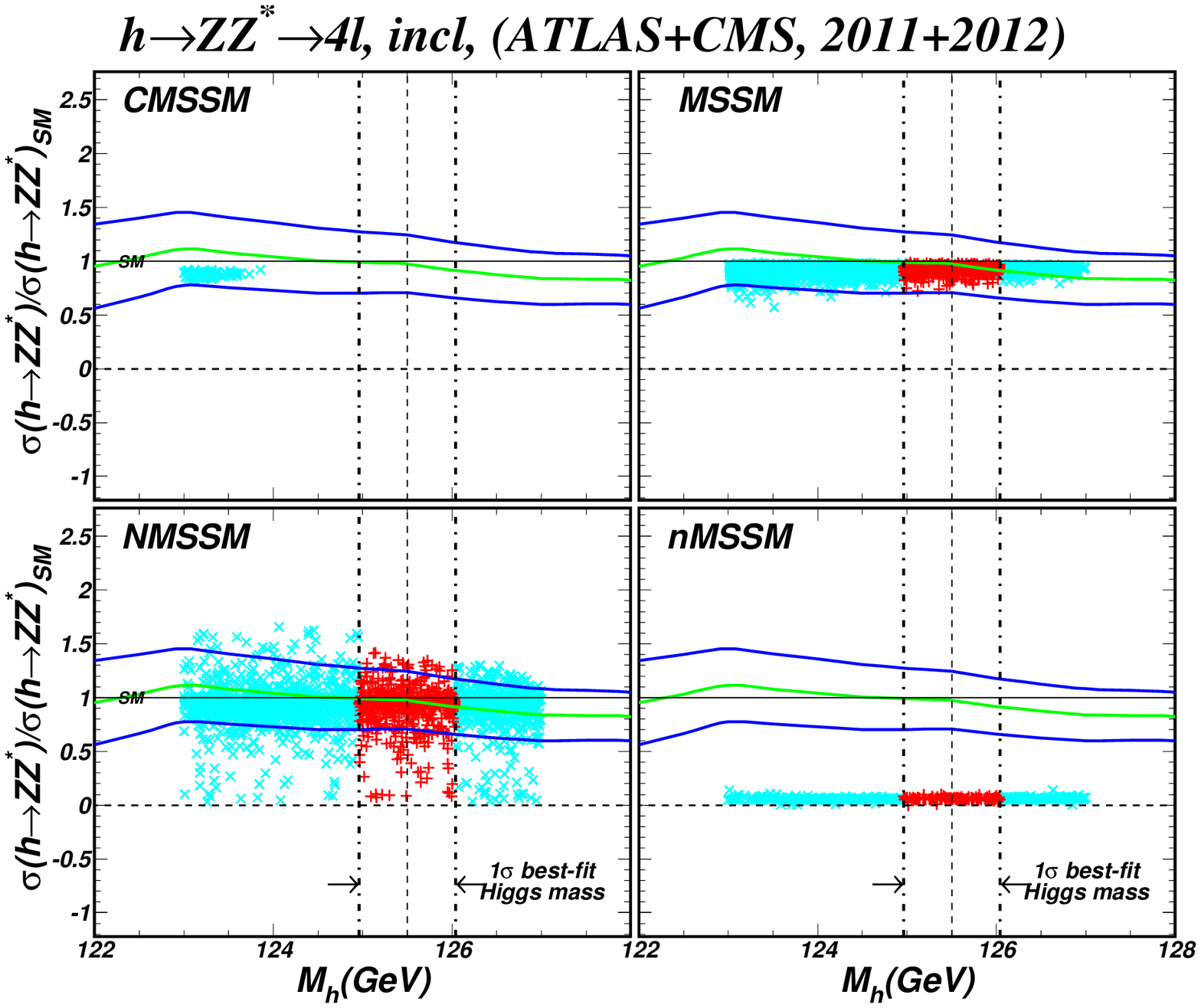}
\vspace*{-0.5cm}
\caption{Same as Fig.1, but showing the signal rate of $pp\to h\to ZZ^* \to 4\ell$
at the LHC.}
\label{fig2}
\end{figure}
In Fig.1 and Fig.2 we project the surviving samples on the planes of
the di-photon rate and the $pp\to h\to ZZ^* \to 4\ell$ rate versus the
SM-like Higgs boson mass. For the experimental curves,
we obtained them by using the method introduced
in \cite{h-fit1} with the ATLAS and CMS data given in \cite{ATLAS,CMS,recent-LHC}.
In combining the data of the two groups, we assume they are independent and Gaussian
distributed. These figures have the following features:
(1) For the CMSSM, $m_h$ is upper bound by about 124 GeV,
significantly away from the $1 \sigma$ best-fit region $125.5 \pm 0.54$GeV \cite{best-fit}.
Considering the theoretical uncertainty of $m_h$ \cite{cao-cmssm},
the maximal value of $m_h$ can be quite close to the best-fit values; but even so
the CMSSM is still disfavored because its di-photon rate is never enhanced.
(2) For the nMSSM, although $m_h$ can easily lie within the $1 \sigma$ best-fit range,
its severely suppressed di-photon and four-lepton rates deviate significantly from
the experimental data (outside the $3\sigma$ range).
(3) For the MSSM and NMSSM, the mass $m_{h}$ and the signal rates of the two channels
can agree with the data at $1 \sigma$ level (as shown later, for the samples in such
  $1 \sigma$ region, the NMSSM is natural while the MSSM needs some extent of
fine-tuning).
(4) Comparing the di-photon data and
the four-lepton data, we see that the former is now more powerful in constraining SUSY.

\begin{figure}[t]
\centering
\includegraphics[width=10.8cm]{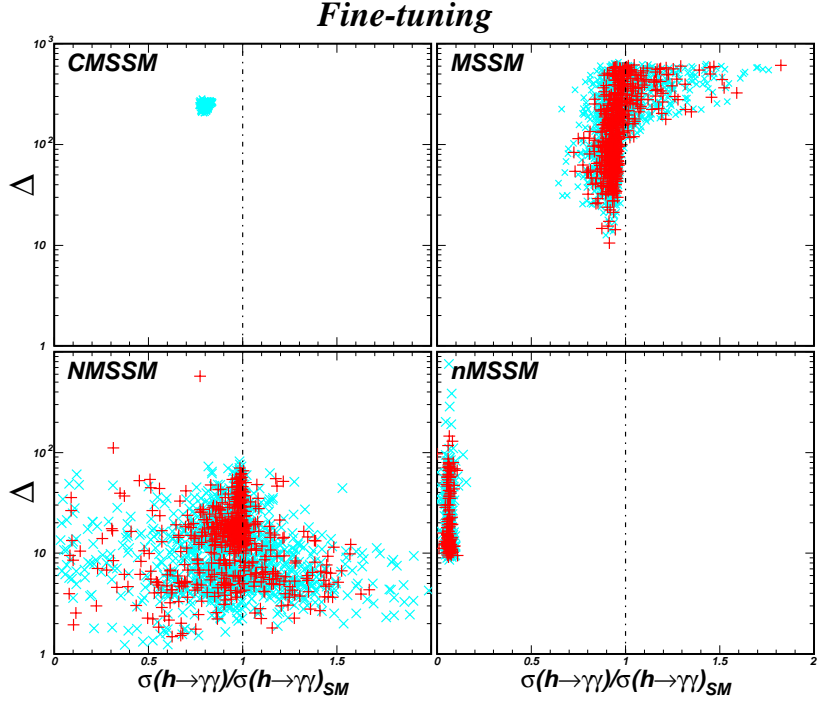}
\vspace*{-0.5cm}
\caption{Same as Fig.1, but showing the fine-tuning extent $\Delta$
versus the di-photon signal rate at the LHC.}
\label{fig3}
\end{figure}
\begin{figure}[t]
\centering
\includegraphics[width=10.0cm]{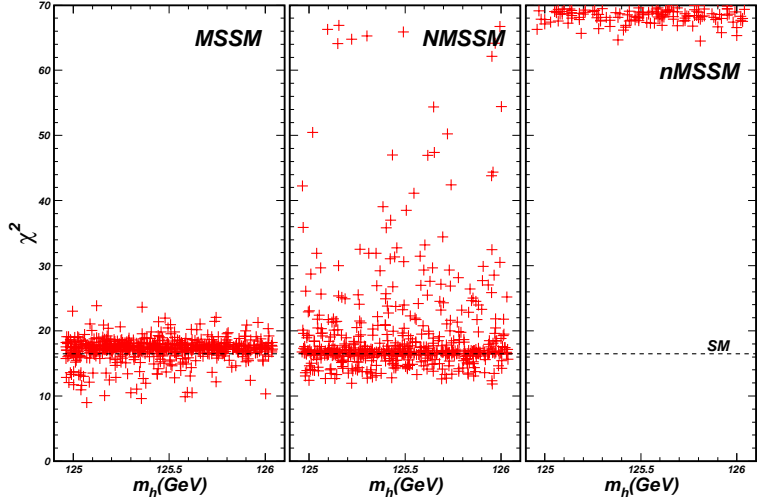}
\vspace*{-0.5cm}
\caption{The golden samples giving $m_h=125.5 \pm 0.54$ GeV,
projected on the plane of $\chi^{2}$ (for 16 degrees of freedom) versus $m_h$.
In calculating $\chi^2$, we took the relevant experimental data from
\cite{recent-LHC,best-fit}. }
\label{fig4}
\end{figure}
\begin{figure}[t]
\centering
\includegraphics[width=12.8cm]{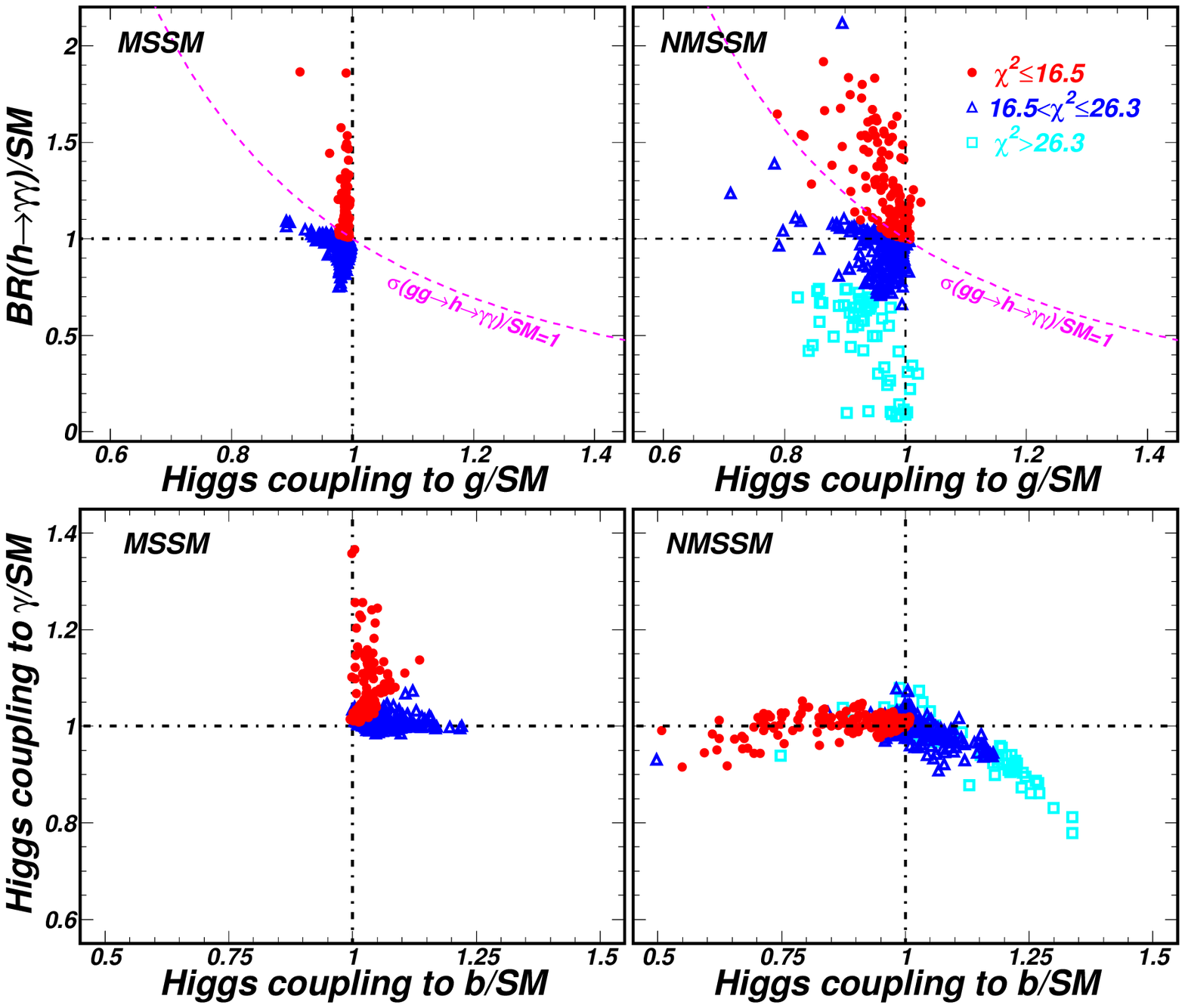}
\vspace*{-0.5cm}
\caption{The surviving samples of the MSSM and NMSSM, showing
the di-photon branching ratio and some couplings of the SM-like Higgs boson.
   Here the surviving samples are classified into three categories
   according to their $\chi^{2}$ values.  The curves in the upper panels
   denote the di-photon signal rate equal to the SM value (the region above
   each curve gives an enhanced di-photon rate at the LHC).}
\label{fig5}
\end{figure}

To further compare the SUSY models, we examine the fine-tuning extent and
 $\chi^{2}$ values for each model.
In Fig.\ref{fig3} we show  the fine tuning
extent $\Delta$ defined in  \cite{ft} versus the di-photon rate.
Same as Fig.1 and Fig.2, only
the samples with $m_h=125\pm 2$ GeV are plotted.
This figure indicates that the
NMSSM with a large $\lambda$ has the lowest tuning extent,
with $\Delta$ as low as 4 for the golden samples with an enhanced
di-photon rate. In contrast,
$\Delta$ in the MSSM is larger than 7 and, in particular,
exceeds 100 for the samples with an enhanced di-photon rate.

For the $\chi^{2}$ values we focus on the  golden samples giving $m_h=125.5 \pm 0.54$ GeV.
In Fig.\ref{fig4} we project these samples on the
plane of $\chi^{2}$ (obtained with 16 degrees of freedom) versus $m_h$.
We compute the $\chi^{2}$ values by the method introduced in \cite{h-fit1}
with the experimental data for $m_h = 125.5 {\rm GeV}$ given in \cite{best-fit}
and in \cite{recent-LHC} (for the latest CMS $\tau^+ \tau^-$ channel).
We assume that the data from different collaborations
and for different search channels are independent of each other.
This figure indicates that the minimal
$\chi^2$ in the MSSM and NMSSM are about $10$, which means that both
models can agree with the LHC data at $1 \sigma$ level, better than the SM.
This figure also indicates that requiring $\chi^2 \leq 26.3$,
which corresponds statistically to the $95\%$ probability for 16 degrees of freedom,
will exclude some samples of the NMSSM and all the samples of the nMSSM.

\begin{figure}[t]
\centering
\includegraphics[width=12.8cm]{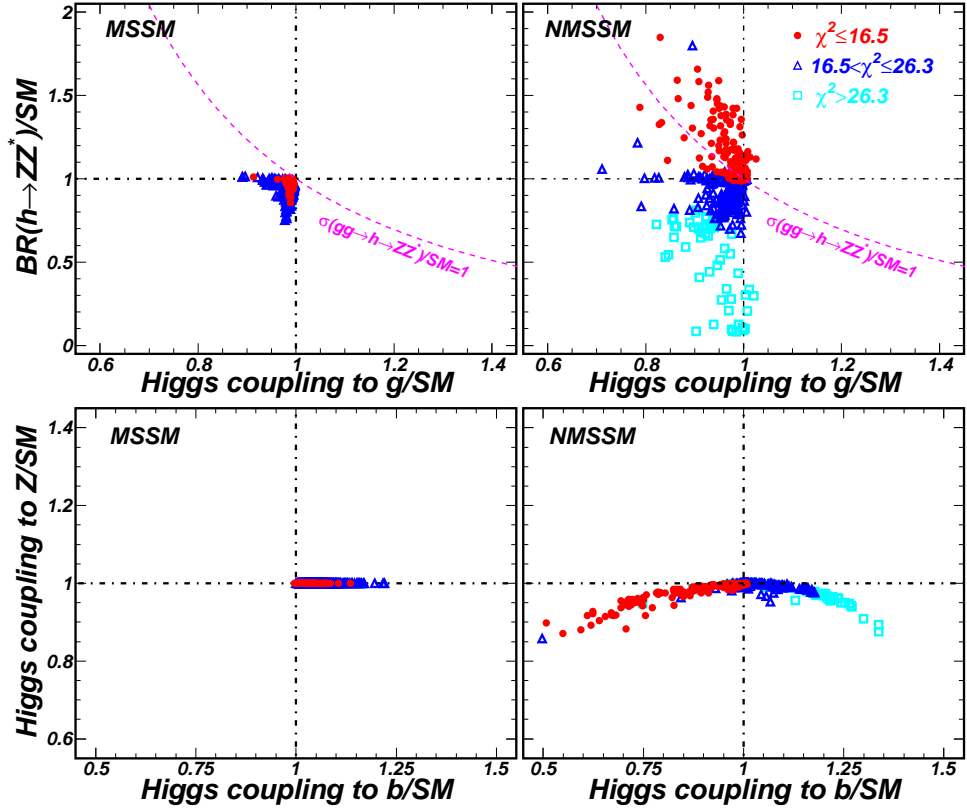}
\vspace*{-0.5cm}
\caption{Same as Fig.\ref{fig5},
   but showing the information of the $4 \ell $ signal at the LHC. }
\label{fig6}
\end{figure}

In our following discussions, we scrutinize the MSSM and NMSSM by
classifying the golden samples into three categories:
$\chi^{2}<16.5$ (better than SM), $16.5<\chi^{2}<26.3$ (worse than SM but in the
$2\sigma$ range) and $\chi^{2}>26.3$ (excluded at $95\%$ CL).
In Fig.\ref{fig5} we show the di-photon branching ratio and some couplings
in the MSSM and NMSSM.
The upper two panels indicate that, in order to get $\chi^2 < 16.5 $,
an enhanced di-photon signal rate is strongly preferred,
which can be realized by a slightly reduced $hgg$ coupling
but a sizably enlarged $h \to \gamma \gamma$ branching ratio.
The two bottom panels indicate that the di-photon branching ratio
is pushed up mainly by the enhanced $h\gamma \gamma$ coupling
(through light $\tilde{\tau}$ loop \cite{carena}) in the MSSM,
and by the suppression of the Higgs width
(through the suppression of the $h b \bar{b}$ coupling) in the NMSSM.
This figure also indicates that in the NMSSM the samples excluded by the Higgs data
are usually characterized by
$\sigma(gg \to h \gamma\gamma)/SM \leq 70\%$,
which is the consequence of the reduced $hgg$ coupling and
the suppressed $Br(h \to \gamma\gamma)$
(through the enlarged $hb\bar{b}$ coupling
or the open-up of new invisible decay $h \to \chi_1^0 \chi_1^0$).
Furthermore, combining Fig.\ref{fig3}
and Fig.\ref{fig5}, one can infer that in the MSSM,
the samples with $\chi^2 < 16.5 $ must
correspond to $\Delta \gtrsim 100 $, which reflects that the
model suffers from some extent of fine tuning to
accommodate the Higgs data;
while for the NMSSM with a large $\lambda$, it is free of such a problem.

\begin{figure}[t]
\centering
\includegraphics[width=12.8cm]{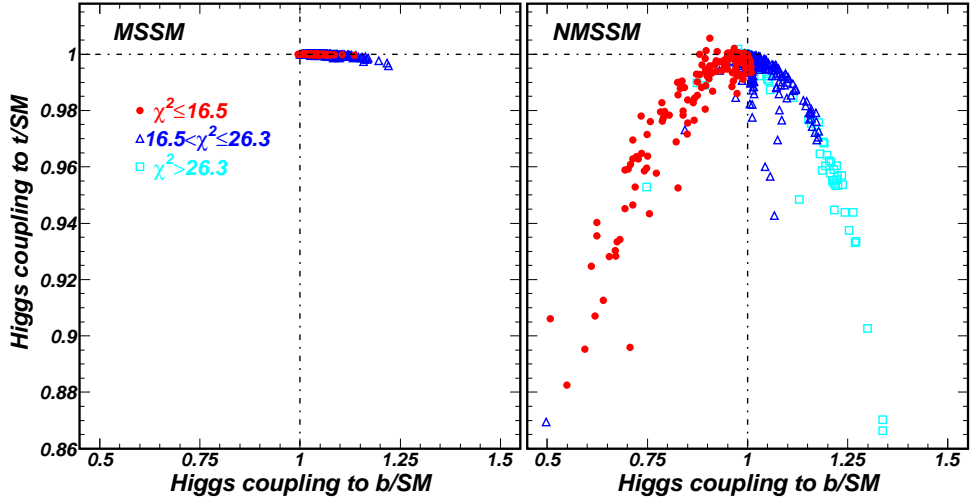}
\vspace*{-0.5cm}
\caption{Same as Fig.\ref{fig5}, but showing the top and bottom quark Yukawa couplings.}
\label{fig7}
\end{figure}

In Fig.\ref{fig6} we show the similar information for the $4 \ell$ signal.
About this figure, three points should be noted. First, unlike the di-photon rate,
an enhanced $4 \ell$ signal rate is unnecessary
to get a low $\chi^2$. In fact, from this figure one can infer  that a slightly reduced
$4 \ell$ signal seems to be more favored.
Second, in contrast with the MSSM where the $4 \ell$ signal
is always reduced, the signal in the NMSSM may be pushed up by the enhancement
of $Br(h \to Z Z^\ast)$ through the reduced $hb\bar{b}$ coupling.
And the last is that although the $hZZ$ coupling in the MSSM  keeps almost the same as
in the SM, $Br(h \to Z Z^\ast)$ in this model is usually smaller than its SM value
due to the enlargement of the Higgs boson width.

\begin{table}
\caption{Detailed information of some samples with low $\chi^2$
in the MSSM and NMSSM.}
  \setlength{\tabcolsep}{2pt}
  \centering
  \begin{tabular}{|c|c|c|c|c|}
    \hline \hline
     & MSSM P1 & MSSM P2 & NMSSM P3 & NMSSM P4 \\
    \hline
     $m_{h}$(GeV) & 125.1 & 125.4 & 125.9 & 125.2
     \\
     $\chi^{2}$ & 9.0 & 9.6 & 11.9 & 12.0
     \\
     $\sigma(h\to \gamma\gamma)/SM$ & 1.59 & 1.82 & 1.45 & 1.35
     \\
     $\sigma(h\to ZZ^{*})/SM$ & 0.86 & 0.98 & 1.21 & 1.16
     \\
     $\Delta$(fine-tuning) & 325.4 & 613.6 & 6.4 & 4.1
     \\
     $\tan \beta$ & 59.9 & 37.1 &  4.7 & 4.0
     \\
     $m_{\tilde{t}_{1}}$(GeV) & 296.6 & 1470.3 & 405.6 & 262.5
     \\
     $m_{\tilde{\tau}_{1}}$(GeV) & 109.7 & 103.4 & 223.4 & 176.2
     \\
     $m_{\tilde{\chi}_{1}^{\rm 0}}$(GeV) & 57.8 & 49.7 & 79.1 & 78.1
     \\
     $\Omega_{CDM}h^2$ & 0.112 & 0.104 & 0.104 & 0.109
     \\
     $Br(B_{s}\to\mu^{+}\mu^{-})/10^{-9}$
      & 5.52 & 9.91 & 3.91 & 3.94
     \\
     $\delta a_\mu/10^{-9}$
     & 3.71 & 2.31 & 0.81 & 0.79
     \\
     $\sigma(hV\to bbV)/SM$
     & 1.01 & 1.00 & 0.62 & 0.73
     \\
     $\sigma(hjj\to WW^{*}jj)/SM$
     & 0.96 & 0.99 & 1.30 & 1.24
     \\
     $\sigma(h\to WW^{*})/SM$
     & 0.86 & 0.98 & 1.21 & 1.16
     \\
     $\sigma(hjj\to \gamma\gamma jj)/SM$
     & 1.77 & 1.85 & 1.57 & 1.45
     \\
     $\sigma(h\to \tau\tau)/SM$
     & 0.86 & 0.99 & 0.55 & 0.64
     \\
     $\sigma^{SI}/10^{-46}$ $({\rm cm}^{2})$
     & 0.32 & 0.04 & 14.3 & 17.0
     \\
  \hline \hline
  \end{tabular}
\label{tab1}\vspace{-0.35cm}
\end{table}

In Fig.\ref{fig7} we show the top and bottom
quark Yukawa couplings. This figure indicates that  the top quark Yukawa
coupling in the MSSM is close to the SM value, while in the NMSSM
it may be suppressed by a factor of 0.85 compared with the SM value.
For the bottom Yukawa coupling, it tends to be enhanced (maximally by a factor of 1.25)
in the MSSM, while in the NMSSM it can be suppressed or enhanced (by a factor
0.5 to 1.3).

\begin{figure}[t]
\centering
\includegraphics[width=13.8cm]{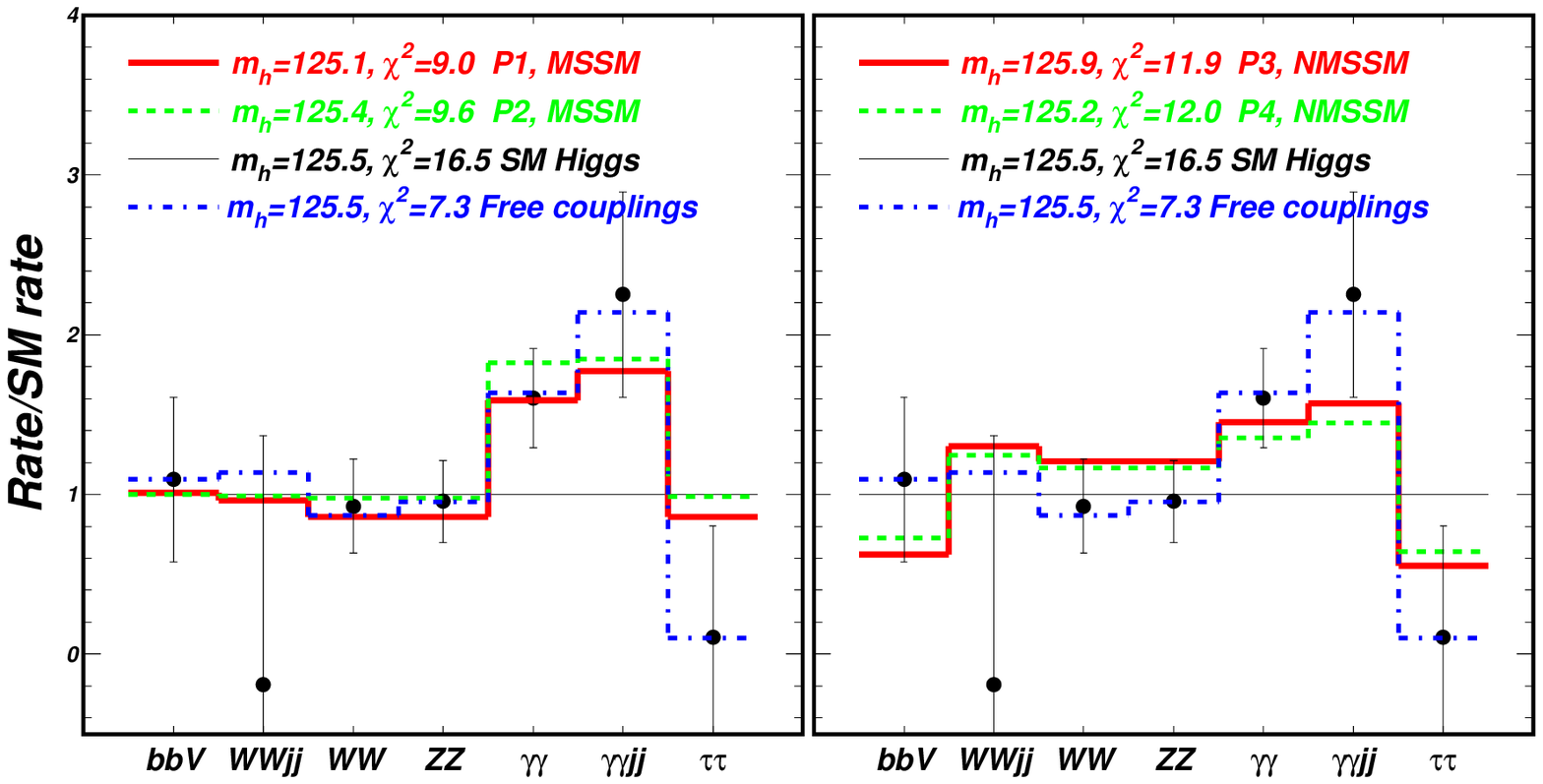}
\vspace*{-0.5cm}
\caption{Predictions for various signal rates of the SM-like Higgs boson
at the LHC. Here two representative samples in the MSSM and NMSSM are displayed
in comparison with the SM prediction and the best-fit values with free couplings.}
\label{fig8}
\end{figure}

In Table \ref{tab1} we present the detailed information
about two representative low-$\chi^2$ samples
for the MSSM and the NMSSM respectively. In order to illustrate how well these samples
are compatible with the experiment data, in Fig.\ref{fig8} we
compare various signal rates with the experimental values given in \cite{best-fit}.
As a comparison, we also show the best-fit rates obtained by varying freely
all the Higgs couplings, including the couplings with photons and gluons
(free coupling scenario). We see that the rates predicted
by these samples agree with the data at $1\sigma$ level except for the channel
$pp \to \gamma \gamma jj$.
We should remind that although the samples in the MSSM
may haver lower $\chi^2$ than the NMSSM with a large $\lambda$,
it is suffered from the fine tuning problem.

\begin{figure}[t]
\centering
\includegraphics[width=13.8cm]{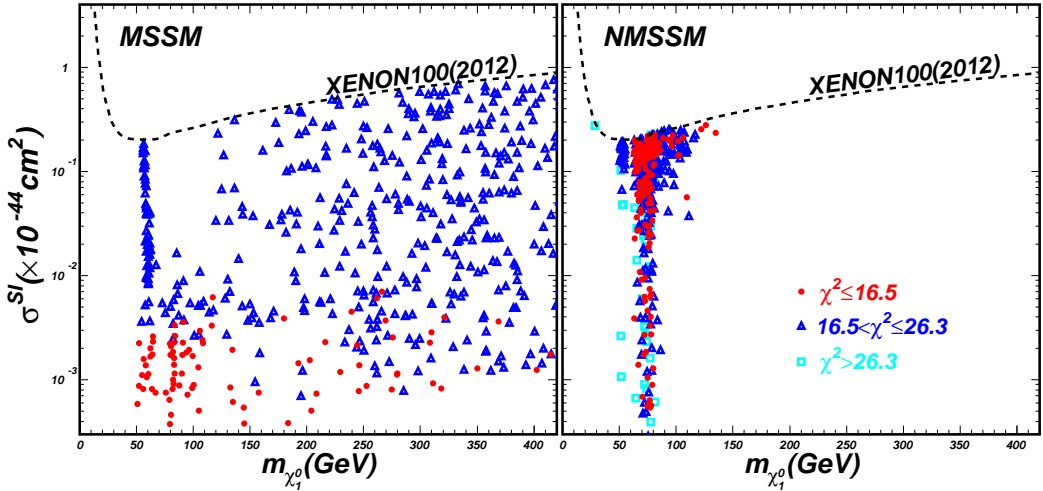}
\vspace*{-0.5cm}
\caption{Same as Fig.\ref{fig5},
but showing the spin-independent scattering cross section
of the neutralino dark matter and the nucleon.}
\label{fig9}
\end{figure}

Finally, we study the direct detection of the neutralino dark matter.
In Fig.\ref{fig9},
we display the spin-independent neutralino-nucleon scattering cross section.
We find that, although the recently released XENON100(2012) data \cite{XENON2012}
have excluded a large portion of the golden samples, there still remain some samples
with the cross section as low as $10^{-46} {\rm cm}^2$, both for the MSSM and NMSSM.
Interestingly, we also find that for the samples with $\chi^2 \leq 16.5$,
the allowed mass of the dark matter is tightly restricted,
varying from 60 GeV to 140 GeV in the NMSSM with a large $\lambda$.

\section{Conclusion}
In this note we compared the properties of the SM-like Higgs boson
 predicted by the low energy SUSY models with
the latest LHC Higgs search data.
For a SM-like Higgs boson around 125 GeV, we obtained the
following observations:
(i) For the MSSM, although it can fit the LHC data quite well,
it is suffered from the fine-tuning problem;
(ii) The most favored model is the NMSSM,
whose predictions can naturally agree with the experimental data at $1\sigma$ level;
(iii) The nMSSM is excluded at $3\sigma$ level due to the much suppressed
di-photon or four-lepton signal rate;
(iv) The CMSSM is quite disfavored
since it is hard to predict a 125 GeV Higgs boson and at same time
cannot enhance the di-photon signal rate.

\section*{Acknowledgement}
This work was supported in part by the National Natural
Science Foundation of China (NNSFC) under grant Nos. 10821504,
11135003, 10775039, 11075045, by Specialized Research Fund for
the Doctoral Program of Higher Education with grant No. 20104104110001,
and  by the Project of Knowledge
Innovation Program (PKIP) of Chinese Academy of Sciences under grant
No. KJCX2.YW.W10.

\end{document}